# Long-Term Cycles in the History of Life: Periodic Biodiversity in the Paleobiology Database

Adrian L. Melott*

Department of Physics & Astronomy, University of Kansas, Lawrence, Kansas, United States of America

### Abstract

Time series analysis of fossil biodiversity of marine invertebrates in the Paleobiology Database (PBDB) shows a significant periodicity at approximately 63 My, in agreement with previous analyses based on the Sepkoski database. I discuss how this result did not appear in a previous analysis of the PBDB. The existence of the 63 My periodicity, despite very different treatment of systematic error in both PBDB and Sepkoski databases strongly argues for consideration of its reality in the fossil record. Cross-spectral analysis of the two datasets finds that a 62 My periodicity coincides in phase by 1.6 My, equivalent to better than the errors in either measurement. Consequently, the two data sets not only contain the same strong periodicity, but its peaks and valleys closely correspond in time. Two other spectral peaks appear in the PBDB analysis, but appear to be artifacts associated with detrending and with the increased interval length. Sampling-standardization procedures implemented by the PBDB collaboration suggest that the signal is not an artifact of sampling bias. Further work should focus on finding the cause of the 62 My periodicity.

Funding: The University of Kansas had no role in study design, data collection and analysis, decision to publish, or preparation of the manuscript.

Competing Interests: The author has declared that no competing interests exist.

* E-mail: melott@ku.edu

## Introduction

The first high significance detection of long-term periodicity in the fossil record is fairly recent [1], based on marine fossil biodiversity over $\sim 500$ $My$. A significant $62\pm 3$ My periodicity was superimposed on the long-term trend, confirmed by a variety of re-analyses of the same data [2–3]. No particular causal mechanism was proposed, but the result was initially published based on its relatively high statistical significance (p = 0.01) and potentially strong implications.

However, these studies were all based on a large compendium [4] which was not controlled for systematic errors such as sampling rate. However, such systematic errors may compromise quantitative study [5,6]. For this reason, an intensive effort (the Paleobiology database: http://paleodb.org) has resulted in a new data set [7,8], constructed, weighted, and subsampled with the intention of minimizing such errors. A statistical study of this dataset concluded with the statement that evidence for autocorrelation did not exist, which result is inconsistent with periodicity [8]. I have extended the analysis around these questions, and found evidence of autocorrelation. I have also found a specific periodic signal consistent with reports based on older data [1–3].

The question of periodicities in fossil biodiversity, or sometimes only in the timings of mass extinction has generated considerable past interest, debate, and discussion. Review of this history is outside the scope of this paper, and can be found elsewhere [3,9]. A few comments are in order: biodiversity periodicity does not depend solely on precise timing for mass extinctions. Any time series can be decomposed into a sum of sinusoids; the question is whether any particular frequencies stand out above the rest. If so, they imply at least a partially repeating pattern. Standard methods are derived from Fourier analysis.

## Methods

In order to do Fourier analysis, long-term trends should be removed: in this case it would be the overall patterns of growth in biodiversity over the last half-billion years. My methods begin by least-squares fit to a cubic of the new, controlled data kindly provided by J. Alroy [7,8]. This is the sample-standardized number of marine invertebrate genera, as published in [7] Figure 1. I have not made any cuts in the data. The cubic is the best-fit of the various simple alternatives, highlighting the general increase over time, with a pause centered around 300 $Ma$. Fits tested were linear, quadratic, cubic, exponential, logarithmic, power, and hyperbolic. Cubic has the highest Coefficient of Determination, either in an absolute sense or adjusted for the number of degrees of freedom. My interest is in whether any repeating patterns of fluctuation about the long-term trend exist. This cubic is less inflected than that shown in [1]; this behavior is a result of the sampling standardization [7], and constitutes part of their new interesting results. Due to constraints of the sampling standardization culling, the temporal intervals are larger than in [1–3]. The temporal bins in this sample based on combining geological intervals are also of unequal length. Their increased size and irregularity are an issue for the analysis, discussed later.

My analysis has been done two ways: 1.) based on the data taken as a function of the intervals (and their midpoints), and 2.) on a file constructed by assigning those values to the time of the mid-point of the interval, and then linearly interpolating between them to assign values every 1 My. A time series running 5–520 *My* spaced at 1 *My* intervals was constructed, using the data [7]. The period 0–5 *Ma* was not included in [7] since the authors argued that its preservation characteristics are very different, due to being close to the present. Interpolation is a well-understood procedure within the context of this method: linear interpolation effectively introduces a smoothing window which reduces amplitudes for high frequencies. My analysis is restricted to frequencies low compared with the smoothing window; the (entirely computable) effect is power lowering amplitude of order 50% at the highest frequency of interest [11–13], and a much smaller effect elsewhere in the range.

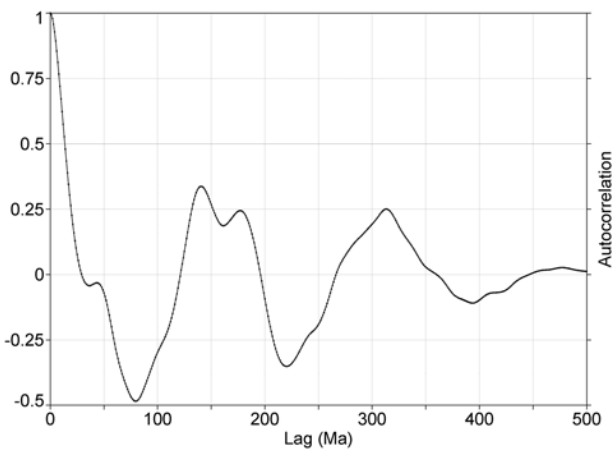

**Figure 1. The autocorrelation of detrended fossil biodiversity, normalized against its value at zero lag, as a function of time.** Note that there is an alternating pattern of peaks and troughs with a period of about 150 My, and extending with declining amplitude to the entire sample interval. The question of periodicity will be treated more quantitatively using power spectra.
doi:10.1371/journal.pone.0004044.g001

Reanalysis of the power spectrum of the data using alternate methods based on the Lomb-Scargle [11–13] transform, which does not require evenly spaced data and therefore no rebinning, is a robust test of whether binning and interpolation introduced any artifacts. In this case the biodiversity number associated with the bin in the data published [7] was used. I performed the analysis both ways. The results are visually identical. The ~50% reduction in power at the highest frequencies, less at lower frequencies, is visible, but significance fits are basically the same since they are based on the height of bumps relative to the spectral mean and trend. Data analysis was performed using AutoSignal 1.7. (http://www.systat.com/products/AutoSignal/). I first note the Hurst exponent [11] $H = 0.92$ over about 100 Myr indicates long-term memory, and excludes white noise which has $H = 0.5$. Power spectral and correlation analysis is designed to investigate what is implied by this initial diagnostic.

## Results

The autocorrelation function can be used to investigate long-term behavior when plotted as a function of time. The time series was extended with zeroes, as needed to prevent a spurious "wraparound" effect [11]. A striking damped-oscillatory pattern can be seen in Figure 1, typically found in natural phenomena that have a strong periodicity. It is strongly suggestive of repetitive behavior on a period of about 150 My. An analysis with correlation based on lag in intervals without time shows a similar shape. Minor changes of shape are introduced, since the variation in interval length mixes temporal frequencies. Data correlated with itself at some time lag is a necessary condition for periodicity to hold [8]. The autocorrelation of biodiversity was plotted only out to a lag of 10 intervals in [8], corresponding to about 110 My. This prevents seeing the full pattern, so [8] concluded that no autocorrelation exists, which would make periodicity impossible. When the full range is plotted, it can be seen as in Figure 1.

Correlation analysis is not the best technique to detect periodicities because the value at any particular lag is a sum over all the oscillations in the data, at different frequencies. In this case a particular frequency signal can be detected clearly because it dominates. Power spectral analysis is to be preferred, since it separates out various frequencies present [11,13].

In order to demonstrate its robustness, I computed power spectrum in two ways. The first is based on the interpolated data as described above, and uses conventional Fast Fourier Transform methods on interpolated data. The spectrum shown on log-log axes in Fig. 2 shows two peaks close to those found previously [1–3], and a new one at higher frequency. The lowest frequency peak has much higher amplitude than seen initially, when it was not significant [1]. The parallel lines correspond to significance levels of 0.05, 0.01, and 0.001 against such a peak appearing anywhere in the spectrum, against an AR(1) fit [11] to the power spectral slope, corresponding to a "red noise" spectrum, as often found in natural time series.

The same analysis was repeated with Lomb-Scargle, with no binning or interpolation, on the data as provided [7], and confirmed that binning and interpolation has not strongly affected the results. The spectrum had slightly increased amplitude with higher frequency, as expected [11–13], including the peaks shown in Figure 2. The same agreement of such different methods was found with different data in earlier work [2,3], and in fact was the main point of [2].

The frequency (f) range shown was restricted in Figure 2 to take account of the limitations due to the length of the total time period and the size of the intervals in the data. The sample used has an average interval length of 11 *My*, and a maximum interval close to 20 *My*. Although Lomb-Scargle may produce results at higher frequencies, based on the Nyquist sampling theorem, reliability degrades around a period $T = 1/f_N \leq 22$ *My*. Results from LS possible at higher frequencies are based on sampling only the short intervals, and are not appropriate when it is not known whether a signal is time-translation independent. This would be a poor assumption for the fossil record. I have plotted down to 25 *My* for informational purposes.

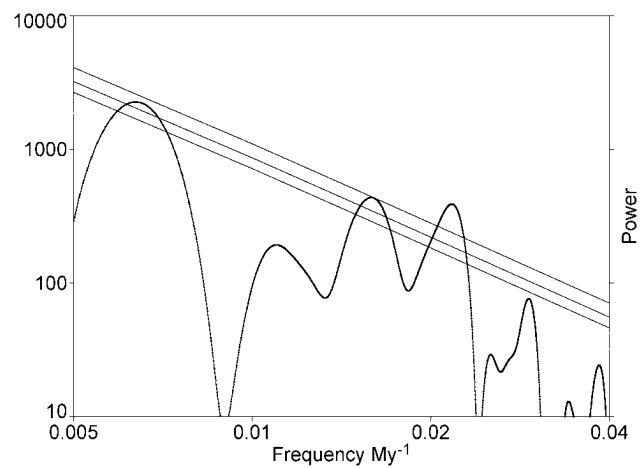

**Figure 2. A logarithmic plot of the power spectrum of fluctuations (determined by Fast Fourier Transform) against frequency in $My^{-1}$.** Higher frequency fluctuations are not shown due to sampling limitations (too close to the interval timescale). The total power is dominated by the area under a few high peaks which exceed confidence limits. These peaks are, from left to right period T = 1/f 157 My, 63 My, and 46 My. Fluctuations outside the plotted frequency range are not shown due to sample limitations (interval length and overall time range). The parallel lines indicate significance at levels p = 0.05, 0.01, and 0.001 against the probability of any such peak arising against the spectral background. Equivalent peaks appear in an analysis based on Lomb-Scargle methods, which do not require binning and interpolation.
doi:10.1371/journal.pone.0004044.g002

The biggest peak is at the lowest frequency, corresponding to a period of 157 (+24 −20) *My*. This provides about half the total variance and is significant at $p = 0.01$ (the probability of a peak so rising above the trend anywhere in the spectrum). It is consistent with a previously detected [1] low significance peak at 140 My. It is a very low frequency, with time for only 3 full periods in the Phanerozoic, but this is taken into account in assigning significance. (Note the negative slope of the fitting lines, so that longer-period power has to satisfy more stringent criteria for significance.) However, as I show later, this result can be questioned. This frequency drives the oscillations shown in Figure 1.

The peak at $f = 0.0158$ closely corresponds to a period of $63.1 \pm 6$ *My*, equal within the (full width half maximum) errors to the earlier [1–3] result $62 \pm 3$ *My*, now at a higher confidence level than formerly, $p = 0.001$.

Another peak at $T = 46$ *My* rises to similar high significance, $p < 0.001$. It is, however, not fully resolved from the 63 My peak, and does not appear in analyses of the Sepkoski compendium [1–3] which has better temporal resolution. The 46 My and 63 My signals together may represent a single peak, possibly resulting from time resolution problems, supported as follows: The interval lengths vary. I have examined their variation by constructing a file of interval lengths in this data as a function of their mid-points. This reveals a very strong spectral feature at 39 *My* (not shown), which indicates alternation of shorter and longer intervals with a 39 My period. A beat between the between the 63 *My* signal in biodiversity and this 39 *My* sampling variation may have produced a large sideband at the mean frequency, in this case corresponding to a predicted period of 48 *My*, close to the new feature. Future versions of the Paleobiology Database with finer temporal resolution will be needed to resolve this issue. In the meantime the 46 and 63 *My* peaks may be regarded as one signal, producing about 20% of the variance in the data set after cubic detrending.

Two peaks close to the first two of the above actually appear in [8], Supporting Information, Fig S2C. A median power method was used there for assigning significance. 95% confidence intervals are plotted, and the non-negative nature of power implies that the axes are logarithmic. I have estimated from the plot that the same peaks rise to about two times and four times the power of the 95% confidence level, making the significance actually higher than the ones I have found. The peaks would surpass lines for substantially higher confidence levels. They are dismissed by assertion as coincidence.

The appearance of long-period spectral peaks in a completely different sample from their original appearance, prepared under controlled conditions, lends support to the reality of the biodiversity variation. This increases the probability that periodicities in biodiversity have existed which are not fossil sampling artifacts, further motivating the search for causal agents which have strongly contributed to the rise and fall of biodiversity on Earth. For this reason additional statistical tests should be applied, if possible across the two data sets.

A combined analysis, additionally using the Sepkoski database as downloaded from Supplementary Information in [1] was done by a generalization of the power spectrum, called the cross-spectrum [11,13] of the two detrended series. The power spectrum of a single time series is essentially $A_i * A_i$, where $A_i$ denotes elements of a series of complex Fourier coefficients as a function of frequency. The power spectrum is therefore real. The cross-spectrum involves the coefficients of two different series: $B_i * C_i$, and it is complex. The amplitude of this complex number is a measure of the extent to which a given frequency $i$ is present in both series; its phase is a measure of the extent to which the cycles in the two series rise and fall in phase with one another. If they are in perfect phase, it is real. I have computed this using my own code, supplemented by IMSL Fourier transforms, as AutoSignal does not have cross-spectral capability. I used the time period 5–505 Ma, common to both data sets. I divided each detrended time series by its own standard deviation, in order to put them on an equal footing.

I show Real($C_{sp}$) the real part of the cross-spectrum, as a function of frequency (Fig 3). Peaks found in Figure 2 are also seen here. (157 to 156 and 63 to 62 are changes that occur due to a combined analysis of two data sets, and are well within the errors associated with either.) Peaks at 156 and 47 My have lower amplitude than at 62 My in Figure 3. This is because, at these places, the function $C_{sp}$ has substantial imaginary component. Even if a given frequency is not a peak in both data sets, it is present and the phase angle can be compared also. This coefficient in the two data sets is out of phase by 1.34 radian at 156 My and by 0.68 radian at 47 My (these correspond to 33 My and 5 My, respectively). Consequently the objective origin of these frequencies are questionable, because although both periodicities appear in the cross-spectrum, the maxima and minima of the cyles do not happen at the same times in the two data sets. Any conclusions about these must be regarded as very tentative, because of this mismatch. If they originated in actual changes in biodiversity, the phases should have good agreement. They may be affected by boundary conditions in one case and temporal resolution in the other.

If I detrend using a linear function, the minimum necessary for detrending, the 157 My signal drops far below the level of significance in both datasets, while the 62 My signal does not. In general the 157 My signal greatly changes its level with various choices of detrending function, while the 62 My does not. Possibly the cubic has in some sense generated the 157 My signal, perhaps interacting with data corrections. The shape of a cubic has non-negligible low frequency power. More tests of this conjecture will appear elsewhere.

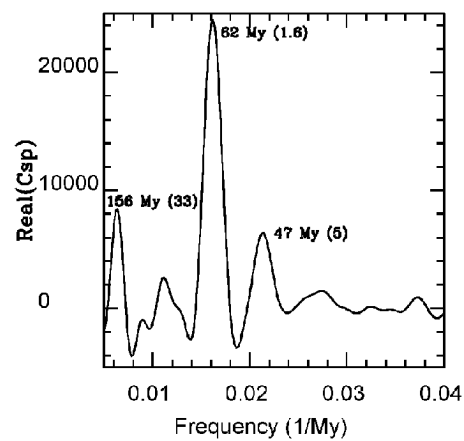

**Figure 3. The real part of the complex cross-spectrum of the Sepkoski data with the Paleobiology Database.** This is a measure of the combined significance of the same frequency in both datasets, with the same phase angles, so that the peaks coincide. The inset numbers give the period corresponding to the shown frequency peak. The number in parentheses is the mismatch, in My, between the peaks in one set versus the other. The 62 My cycle dominates the figure and has excellent phase agreement between the two compendia.
doi:10.1371/journal.pone.0004044.g003

Contrarily, the cross-spectral peak at 61.7 (+4 −3, FWHM) My is completely robust. Its phase displacement is only 0.16 radian between the two sets, corresponding to a 1.6 My difference in the placement of peaks and valleys of a much longer cycle. The two data sets do share some common data, but there is also substantial additional data in PBDB, and the treatment of the data has been completely differerent [4,7]. In particular, strong work toward sampling-standardization [7] should effectively remove much bias based on temporal variation in the quality of the fossil record. The 62 My periodicity appears in two largely independently generated data sets, with multiple methods of analysis.

## Discussion

I have shown that a periodicity at 62±3 My with essentially identical period and phase which was uncovered [1] in the Sepkoski dataset [4] also appears in the Paleobiology Database data [7]. It appears in both interpolation methods and Fast Fourier Transform, or Lomb-Scargle methods, which do not require this. Since PBDB have exercised strong control over biases, particularly those related to sampling rates, this suggests (but of course does not prove) that the periodicity is not a consequence of some sampling bias. Further discussion should emphasize possible causal mechanisms.

While the point of this paper is not to demonstrate any causal mechanism, I summarize those suggested to date that either (a) have a theoretical reason which has been argued to produce such a periodicity, or (b) have empirically demonstrated some closely related periodicity.

One possibility is that there is simply a long delay for recovery from extinction events [14]. If there were a characteristic recovery timescale of order 62 My, and random events lowering biodiversity, this could easily produce a periodicity such as observed.

One causal clue is an observed strong correlation between the 62 My biodiversity cycle and $^{87}Sr/^{86}Sr$ isotope ratio [15], commonly used as a proxy for the erosion rate of continental crust, and was suggested in the context of uplift, particularly that episodes of volcanic activity were associated with lower biodiversity. The reported isotopic signal is strong, but a theoretical basis for assigning uplift a 62 My period with little variation throughout the Phanerozoic is absent. If uplift were demonstrated to be periodic, this would be a strong argument. A lowered biodiversity from any cause including reduced plant cover might increase erosion rates, producing the isotopic signal.

A second independent result at low significance is a periodic signal at 61 My, $p = 0.14$ in the evolution of new gene families subsequent to gene duplication [17]. Major fluctuations were observed to follow closely timings of lowered biodiversity.

Thirdly, biodiversity declines correspond in timing with excursions of the Solar System to Galactic north, and they are possibly caused by the effects of a resulting increased exposure high-energy cosmic rays ($TeV$ to $PeV$) [18,19]. The large literature on cosmic rays and cloud formation is consistent with the data that such irradiation could affect long-term trends in cloud formation [20], although cosmic ray variability probably cannot explain *recent* climate change [21]. An additional effect of enhanced cosmic rays would be a major increase of muons on the ground (and in the sea, since they easily penetrate 1 km of water). Muons are responsible for about half the present penetrating radiation dose in North America [22], and a large increase would damage organisms [23–25], providing a long-term stress. Some stress may also result from increased exposure to Solar ultraviolet-B radiation penetrating the atmospheric ozone layer, depleted by chemical changes from the ionizing cosmic rays [19]. Long-term stress can increase the severity of biodiversity declines due to impulsive events according to [26].

Sea level changes are strongly correlated with fossil biodiversity changes from the late Triassic to the Pliocene [27]. However, while this account describes a correlation, it does *not* report a sea level cyclicity, leaving open the possibility that sea level is a contributing component to fossil biodiversity changes, but not the one that drives the 62 My signal. In fact, the part of the biodiversity data primarily driving the 62 My signal has been reported to come from the Jurassic and earlier [3], having little overlap with the period studied in [27]. Also, more recent analysis [28] finds that sea-level changes are a possible biodiversity driver. However, [28] concludes that they fit the data better if they function as a causal agent and not a sampling bias. This is consistent with my finding that the 62 My biodiversity signal arises in the PBDB data, which has had a strong sampling bias correction, and is also consistent with the Sr isotope result [15]. Further work should include a careful examination of modern sea level data prior to 150 Mya, including a cross-spectral comparison with biodiversity data.

I know of no other mechanisms which would be expected to produce a 62 My periodicity, or other coincident data which are in phase with the biodiversity fluctuations. This, of course, does not mean that they may not exist.

## Acknowledgments

I thank J. Alroy for providing the data, and R. Bambach, M. Benton, B. Lieberman, B. Thomas, and referee J. Lipps for useful comments which improved the manuscript.

## Author Contributions

Analyzed the data: ALM. Wrote the paper: ALM.